\documentclass[%
 reprint,
 amsmath,amssymb,
 aps,
floatfix,
]{revtex4-1}

\usepackage{graphicx}
\usepackage[space]{grffile}
\usepackage{latexsym}
\usepackage{textcomp}
\usepackage{longtable}
\usepackage{multirow,booktabs}
\usepackage{amsfonts,amsmath,amssymb}
\usepackage{natbib}
\usepackage{url}
\usepackage{hyperref}
\hypersetup{colorlinks=false,pdfborder={0 0 0}}
\newif\iflatexml\latexmlfalse
\DeclareGraphicsExtensions{.pdf,.PDF,.png,.PNG,.jpg,.JPG,.jpeg,.JPEG}

\usepackage[utf8]{inputenc}
\usepackage[ngerman,english]{babel}

\begin{document}

\title{Angle-resolved RABBIT: theory and numerics}

\author{Paul Hockett}
\email{paul.hockett@nrc.ca}
\affiliation{National Research Council of Canada, 100 Sussex Drive, Ottawa, K1A
0R6, Canada}

\maketitle

\section{Abstract}
Angle-resolved (AR) RABBIT measurements offer a high information content measurement scheme, due to the presence of multiple, interfering, ionization channels combined with a phase-sensitive observable in the form of angle and time-resolved photoelectron interferograms. In order to explore the characteristics and potentials of AR-RABBIT, a perturbative 2-photon model is developed; based on this model, example AR-RABBIT results are computed for model and real systems, for a range of RABBIT schemes. These results indicate some of the phenomena to be expected in AR-RABBIT measurements, and suggest various applications of the technique in photoionization metrology.

\medskip 
\textit{Article history}

\begin{itemize}
\item arXiv (this version)
\item \href{https://www.authorea.com/users/71114/articles/152997-angle-resolved-rabbit-theory-and-numerics/_show_article}{Original article (Authorea)}

DOI: 10.22541/au.149037518.89916908
\end{itemize}

\textit{See also}
\begin{itemize}
\item \href{https://doi.org/10.6084/m9.figshare.4702804}{AR-RABBIT results presentation}

DOI: 10.6084/m9.figshare.4702804

\item \href{https://doi.org/10.6084/m9.figshare.c.3511731}{Background material on angle-resolved photoionization (and refs. therein)}

DOI: 10.6084/m9.figshare.c.3511731
\end{itemize}

\section{Introduction}

\begin{figure}[h!]
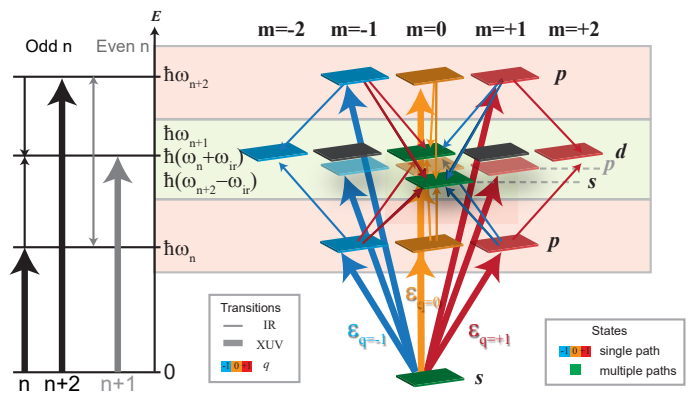

\begin{center}
\includegraphics[width=1.1\columnwidth]{{{ionization_paths_incE_080217}}}
\caption{{\label{fig:pathways}Energy and angular-momentum state diagram for RABBIT processes, starting from an initial $s$-state. The upper and lower panels show states accessed by absorption of a single XUV photon, of harmonic order $n$ or $n+2$, and field polarization $q$. The middle panels shows states accessed by 2-photon pathways, involving subsequent absorption or emission of an IR photon, hence corresponding to a usual RABBIT sideband. States are colour-coded according to contributing pathways. For schemes involving even harmonics, absorption of harmonic $n+1$ results in additional accessible states, at the same energy as the usual sidebands, but different angular momentum - these are the $p$-states in the middle panel.%
}}
\end{center}
\end{figure}

The RABBIT methodology - ``reconstruction of attosecond harmonic beating by interference of two-photon transitions" \cite{Muller_2002} - essentially defines a scheme in which  XUV pulses are combined with an IR field, and the two fields are applied to a target gas. The gas is ionized, and the photoelectrons detected. In the typical case, the IR field is at the same fundamental frequency $\omega$ as the field used to drive harmonic generation, and the XUV field generated is an atto-second pulse train with harmonic components $n\omega$, with odd-$n$ only. In this case, if the intensity of the IR field is low to moderate, the resultant photoelectron spectrum will be comprised of discrete bands corresponding to direct 1-photon XUV ionization, and sidebands corresponding to 2-photon XUV+IR transitions \cite{Muller_2002}. (The energetics of this situation are illustrated in fig. \ref{fig:pathways}.)  Temporally, if the XUV pulses are short relative to the IR field cycle, the sidebands will also show significant time-dependence, since they will be sensitive to the optical phase difference between the XUV and IR fields, with an oscillatory frequency of $2\omega$. In this case, a measurement which is angle-integrated, or made at a single detection geometry, can be viewed as a means to characterising the properties of the XUV pulses (spectral content and optical phase), provided that the ionizing system is simple or otherwise well-characterised \cite{Muller_2002}; RABBIT can therefore be utilised as a pulse metrology technique \cite{Muller_2002,Krausz_2009}, and this is the typical usage.

Conversley, RABBIT can also be regarded as a photoelectron metrology technique, since it is sensitive to the magnitudes and phases of the various photoionization pathways accessed. In contrast to most traditional (energy-resolved) photoelectron spectroscopy techniques, RABBIT has the distinction of interfering pathways resulting from different 1-photon transition energies: it is thus sensitive to the energy-dependence of the photoionization dynamics, as well as to the partial-wave components within each pathway. 
An angle-resolved (AR) RABBIT measurement is particularly powerful in this regard, since the partial-wave phases are encoded in the angular part of the photoelectron interferogram. Although this is a potentially powerful technique, the underlying photoionization dynamics may be extremely complicated, hence quantitative analysis of experimental results is challenging.

In essence, AR-RABBIT can therefore be considered as a technique which combines traditional photoionization and scattering physics with an additional (time-dependent) perturbation in the form of the IR laser field. This field provides additional couplings between the 1-photon (XUV) channels. In the usual RABBIT intensity regime, these two steps can be decoupled, allowing for the XUV absorption to be treated as a weak-field bound-free transition (photoionization), 
followed by absorption of an IR photon - this latter step is a transition purely between different free electron states in the continuum, often termed continuum-continuum coupling. This scheme is illustrated in the energy-domain in fig. \ref{fig:pathways}(left). Therefore, the problem becomes one of dealing with 
a two-photon matrix element, describing these two sequential light-matter interactions. Furthermore, if the continuum-continuum coupling is assumed to be at long-range (i.e. temporally and spatially distinct from the bound-continuum coupling of the first, bound-free, step, and at the asymptotic limit of the continuum wavefunction), then a simplified treatment can be developed for this second transition. In this vein, Dahlstr\selectlanguage{ngerman}ö\selectlanguage{english}m, L'Hullier and coworkers have done significant work, including angle-integrated resonant cases and extensive theoretical treatments of the problem. See, for instance, \textit{Introduction to attosecond delays in photoionization} \cite{Dahlstr_m_2012} and \textit{Study of attosecond delays using perturbation diagrams and exterior complex scaling} \cite{Dahlstr_m_2014} for general background theory and perturbative treatments similar to those discussed herein, \textit{Phase measurement of resonant two-photon ionization in helium} \cite{Swoboda_2010} for a specific example (angle-integrated), and \textit{On the angular dependence of the photoemission time delay in helium} \cite{Ivanov_2016} for work on this specific angle-resolved case. 

In this work, the same basic conceptual path to modelling RABBIT as a sequential two-photon process is followed, but the emphasis is placed on the role of the photoinization dynamics. 
This provides a route to the modelling and analysis of angle-resolved RABBIT, based on canonical photoionization theory and employing a full partial-wave treatment of the continuum. 
Following the similar treatment of ref. \cite{Hockett_2015}, which investigated sequential 3-photon ionization in a time-dependent IR field (conceptually similar to a RABBIT scheme), the electric fields are modelled in a circular basis to allow for arbitrary field polarization states.
The treatment is general, and applicable to any atomic or molecular system, provided that the IR field can be neglected for the first step. Essentially, within this framework angle-resolved RABBIT can be considered as an extension of traditional angle-resolved photoelectron measurements, and many of the same fundamental considerations and potential applications apply \cite{Reid_2003, Reid_2012}. As usual, in cases where the XUV and/or IR field is strong, only full numerical treatments are capable of correctly describing the coupled light-matter system (see, for instance, refs. \cite{Bondar_2009,Spanner_2013,Gal_n_2013}), and this regime is not within the scope of the perturbative model discussed herein.

In the following, a framework for AR-RABBIT modelling is defined in terms of the general form of the required photoionization matrix elements, the final continuum wavefunctions and the resultant observables (sect. \ref{sec:theory}). This framework is then applied to simple model cases (sect. \ref{sec:model-systems}), in order to develop a phenomenological understanding of AR-RABBIT measurements. To explore the application of the framework to real systems (sect. \ref{sec:real}), numerical treatments for the radial matrix elements are detailed (sect. \ref{sec:numerics}), and the framework is applied to model a range of specific AR-RABBIT measurements of neon.

\section{Theory \label{sec:theory}}
In this section, a basic theoretical framework for AR-RABBIT is defined. Further numerical details are discussed in sect. \ref{sec:real}.

\subsection{1-photon ionization by the XUV field}

The dipole matrix element for 1-photon ionization by the XUV field, corresponding to direct ionization from
an initial bound state $|n_{i}l_{i}m_{i}\rangle$ to a final continuum
state $|l_{f}m_{f};\,\mathbf{k}\rangle$, is given as:

\begin{eqnarray}
d_{xuv}(\mathbf{k},\, t) & = & \langle\mathbf{k};\, l_{f}m_{f}|\hat{\boldsymbol{\mu}}_{if}.E(\Omega,\, t,\, q)|n_{i}l_{i}m_{i}\rangle\\
 & = & R_{l_{i}l_{f}}(k)E_{xuv}^{q}(\Omega,\, t)\langle l_{f}m_{f},1q|l_{i}m_{i}\rangle
\end{eqnarray}
where $\boldsymbol{\hat{\mu}}_{if}$ is the dipole operator. In the second line, the matrix element is decomposed in terms of radial and geometric parts. Here $R_{l_{i}l_{f}}(k)$
denotes the radial integrals, which are dependent on the
magnitude of the photoelectron wavevector $\mathbf{k}$; 
$\langle l_{f}m_{f},1q|l_{i}m_{i}\rangle$
is a Clebsch-Gordan coefficient which describes the angular momentum
coupling for single photon absorption, where the field polarization
(circular basis) is defined by $q$, and the spectral ($\Omega$) and temporal ($t$) properties of each polarization component by $E_{xuv}^{q}(\Omega, t)$.

This matrix element is essentially identical to canonical treatments for 1-photon ionization (e.g. Cooper \& Zare \cite{Cooper_1968, Cooper1969} for atomic photoionization, Dill \cite{Dill_1976} for fixed-molecule photoionization), apart from the inclusion of a time-dependent $E$-field. In this decomposition, the Clebsch-Gordan coefficients can be calculated analytically, the $E$-field can be defined analytically or numerically, and the $R_{l_{i}l_{f}}(k)$ (complex) require numerical solution for a given ionizing system.  Essentially, the analytical part of the solution encodes the angular momentum selection rules, while the $R_{l_{i}l_{f}}(k)$ provide the amplitude and phase coefficients for each partial-wave channel for a specific problem (ionizing system and energy). The notation used here implicitly assumes that the radial integrals $R_{l_{i}l_{f}}(k)$ are independent of $m_{i}$ and $m_{f}$. For atomic systems this is a good approximation, and allows for a simplified treatment of the photoionization dynamics, but for molecules this assumption does not hold (due to the loss of spherical symmetry in the core region) and all $m$ components must be treated explicitly (see, e.g., refs. \cite{Dill_1976, Park_1996}).

\subsection{Continuum-continuum coupling}

The transition between two continuum states, $i$ and $f$, further labelled by energy and angular
momentum, coupled by 1-photon absorption or emission from the IR field, can be similarly given as:

\begin{eqnarray}
d_{ir}(\mathbf{k_{i}},\,\mathbf{k_{f}},\, t) & = & \langle\mathbf{k_{f}};\, l_{f}m_{f}|\hat{\boldsymbol{\mu}}_{if}.E(\Omega,\, t,\, q)|\mathbf{k_{i}};\, l_{i}m_{i}\rangle\\
 & = & R_{l_{i}l_{f}}(k_{i},k_{f})E_{ir}^{q}(\Omega,\, t)\langle l_{f}m_{f},1q|l_{i}m_{i}\rangle
\end{eqnarray}

Note that, as for bound-free ionization, the radial part of the matrix elements $R_{l_{i}l_{f}}(k_{i},k_{f})$ are here not defined explicitly, but must be considered specifically for the problem at hand.

\subsection{Final state wavefunctions}

The final continuum states populated are given by expansions in continuum partial-waves $|l_{f}m_{f};\,\mathbf{k}\rangle$. 
The expansion parameters are defined by the matrix elements given above, for the various pathways of interest in a RABBIT scheme, as:
\begin{itemize}
\item One photon (XUV) final states\\
\begin{equation}
\Psi_{xuv}(\mathbf{k},\, t)=\sum_{l_{f}m_{f},l_{i}m_{i}}d_{xuv}(\mathbf{k},\, t)|l_{f}m_{f};\,\mathbf{k}\rangle
\end{equation}

\item Two photon (XUV+IR) final states\\
\begin{equation}
\Psi_{\pm}(\mathbf{k},\, t)=\sum_{l_{f}m_{f},l_{v}m_{v},l_{i}m_{i}}d_{xuv}(\mathbf{k_{v}},\, t)d_{ir}(\mathbf{k_{v}},\,\mathbf{k},\, t)|l_{f}m_{f};\,\mathbf{k}\rangle
\end{equation}
where the $\pm$ refers to absorption or emission of an IR photon,
and $v$ denotes the intermediate 1-photon continuum states.

\item Generic channel summed and partial-wave resolved final states. This notation simply indicates a final state which is the resultant sum over various ionization channels $c$, each decomposed into a set of final $|l_{f}m_{f}\rangle$ waves, and serves as a general short-hand.\\
\begin{equation}
\Psi(\mathbf{k},\, t)=\sum_{c}\Psi_{c}(\mathbf{k},\, t)=\sum_{c}\sum_{l_{f}m_{f}}\psi^c_{l_{f}m_{f}}(\mathbf{k},t)\label{eq:Psi_gen}
\end{equation}

\end{itemize}

In this case, the number of angular momentum components $(l,~m)$ involved depends on the ionizing system. For centro-symmetric systems (e.g. hydrogen), $l$ is a good quantum number and only bound-free transitions with $\Delta l=\pm1$ are allowed; this is also usually a reasonable approximation for multi-electron atomic systems. However, as eluded to previously, for molecular systems many angular momentum components are typically expected, due to the loss in symmetry of the scattering potential at short range, and the problem becomes more complex; for discussion on this topic see, for instance, refs. \cite{Dill_1976,Park_1996}.

In this treatment, $t$ denotes the temporal dependence of the final states, due to both laser fields $E(\Omega,\, t,\, q)$. This dependence can be simplified to a dependence upon only the relative XUV to IR field delay, $\tau$, under the assumption that the XUV field is short relative to the rate of change of the IR field. In the limiting case, the time-dependence of the XUV field is a $\delta$-function, and only the instantaneous properties of the IR field at $t=\tau$ are important, hence no temporal integration is required.

\subsection{Observables}

The energy and angle resolved photoelectron measurements, as a function
of the XUV-IR delay $\tau$, are then given by:
\begin{itemize}
\item One photon transitions, single path - direct ionization, the usual case
for odd-harmonic bands in standard RABBIT experiments (odd-harmonics only in the XUV spectrum). Note this signal is
effectively time-independent, since the signal does not depend on
$\tau$\\
\begin{equation}
I_{1}(E,\theta,\phi)  =  \Psi_{xuv}(\mathbf{k},\, t)\Psi_{xuv}^{*}(\mathbf{k},\, t)
\end{equation}

\item Two photon matrix elements, with two paths - usual RABBIT sidebands for an XUV spectrum with odd-harmonics only\\
\begin{eqnarray}
I_{2}(E,\theta,\phi,\,\tau) & = & (\Psi_{+}(\mathbf{k},\,\tau)+\Psi_{-}(\mathbf{k},\,\tau))\times c.c.
\end{eqnarray}

\item One \& two photon paths - all photoelectron bands for ``extended" RABBIT experiments, when the XUV spectrum also contains even-harmonics\\
\begin{equation}
I_{3}(E,\theta,\phi,\,\tau) = (\Psi_{+}(\mathbf{k},\,\tau)+\Psi_{-}(\mathbf{k},\,\tau)+\Psi_{xuv}(\mathbf{k},\,\tau))\times c.c.
\end{equation}

\end{itemize}

In all cases the resultant observable, for each photoelectron band observed in a RABBIT scheme, centered at photoelectron energy $E$, can be described by an expansion
in spherical harmonics $Y_{L,M}$ with time-dependent expansion parameters $\beta_{L,M; E}(\tau)$, and a Gaussian radial function:

\begin{equation}
I(E,\theta,\phi,\,\tau) = \sum_{L,M}\beta_{L,M; E}(\tau)Y_{L,M}(\theta,\phi)G(E,\sigma)
\end{equation}

In practice, the Gaussian features centered at energies $E$, of width $\sigma$, are defined from the experimental harmonic spectrum. In principle, the energy dependence of the matrix elements across each photoelectron band, and the effect of this dependence on the radial spectrum, should be considered; however, in many cases it is reasonable to assume that the matrix elements are smoothly varying as a function of energy, and may be approximated as constant for each discrete photoelectron band (typically spanning a few 100 meV) - hence a single energy point at the peak of the band is assumed to be representative of the band. This is essentially the ``smoothly varying continuum" (SVC) approximation, but will clearly break-down in the presence of any sharp features such as autoionizing resonances. For more general discussion, in the context of photoionization and the energy-dependence of PADs, see refs. \cite{Park_1996,Seideman_2001}.

\section{Model systems \label{sec:model-systems}}

While general, the preceding treatment does not offer much direct insight since many details remain to be defined - specifically the angular momentum states which play a role, and the radial integrals. In order to proceed, one must model specific cases, thus select appropriate initial states and compute the relevant integrals - for example,  Dahlstr\selectlanguage{ngerman}ö\selectlanguage{english}m et. al. have presented specific results for a hydrogenic treatment, including different levels of approximation \cite{Dahlstr_m_2012}. Herein, the cases of ``standard" and ``extended" RABBIT are explored, starting with a basic model system to provide physical insight, while sect. \ref{sec:real} details specific real cases.

\subsection{Sidebands in standard AR-RABBIT\label{sec:usual-RABBIT}}

The ``usual" RABBIT sidebands result from two interfering pathways, corresponding to 2 photon transitions via $H(n)+IR$ and $H(n+2)-IR$, where $H(n)$ denotes a harmonic of order $n$. The corresponding wavefunctions were denoted by $\Psi_+$ and $\Psi_-$ above. To model this, and explore paradigmatic behaviours, the dipole matrix elements required can be set as model parameters, and the energy dependence of the pathways neglected. This provides a model in which the angular interferograms, and temporal behaviour, can be probed. 

Fig. \ref{fig:pathways} illustrates a basic RABBIT scheme, for the simplest model system. Ionization is from a pure $s$-state, resulting in ionization pathways $s\xrightarrow{xuv} p\xrightarrow{\pm ir} s+d$. To model this case, identical radial matrix elements were set for each 2-photon channel (denoted $c$), with variable phases:
\begin{eqnarray}
R_{s\rightarrow p}^{c} & = & 1e^{i\phi_{s,p}^{c}}\\
R_{p\rightarrow d}^{c} & = & 1e^{i\phi_{p,d}^{c}}\\
R_{p\rightarrow s}^{c} & = & 0.3e^{i\phi_{p,s}^{c}}
\end{eqnarray}

\begin{figure*}[h!]
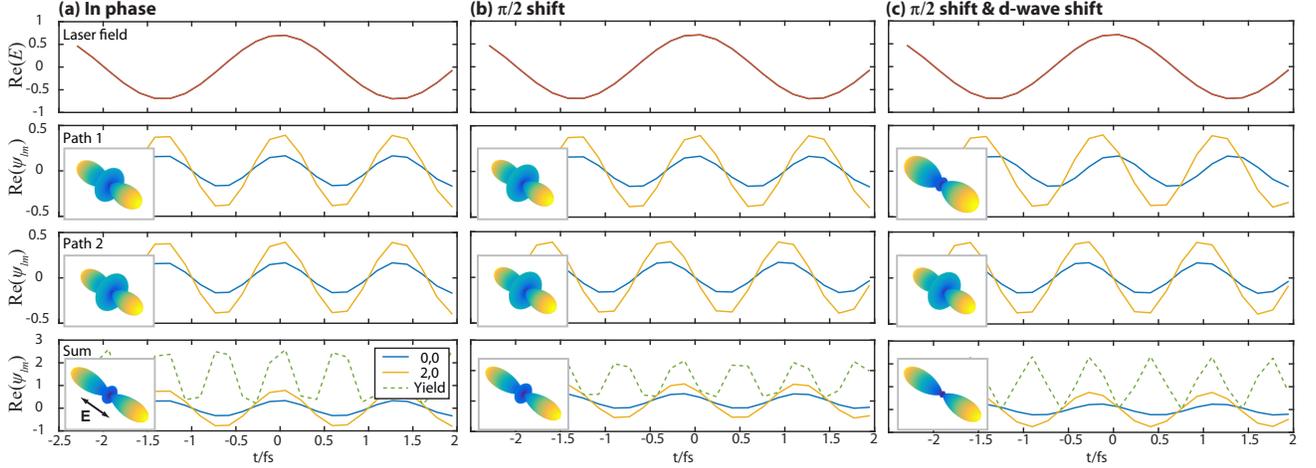

\begin{center}
\includegraphics[width=2\columnwidth]{{{basic_model_3up_m0_only_inc_PADs_080317}}}
\caption{{\label{fig:2-state_model}Partial-wave resolved RABBIT traces. Each column shows, from top to bottom, the IR laser field, the two paths to the side-band and the resultant sum over paths. The plots for each path show the real part of the final-state wavefunction $Re[\psi_{l,m}^c]$ as a function of partial-wave and time, and the sum over paths $Re[\sum_c\psi_{l,m}^c]$, where $c$ is a channel index for path 1 and path 2. The legends show the various $l,m$ partial-waves plotted, and the final (angle-integrated) photoelectron yields $I_2(\tau)$. Insets show the (time-invariant) angular interferograms, $I_2(\theta,\phi)$, for each individual ionization channel, and the resultant sum. Note the slightly reduced scale for the yields plotting in column (b), in this case of partly destructive intereference the temporal modulations are reduced.%
}}
\end{center}
\end{figure*}

Fig. \ref{fig:2-state_model} shows the results for these case, in which the laser fields are set to $q=0$ only (linear polarization), and the XUV phases are set to zero. The phases of the dipole matrix elements were varied to probe the behaviour of the sidebands, and the three example cases have the following phases set:

\begin{description}
\item[(a)] All phases set to 0.
\item[(b)] $\phi_{s,p}^{2}=\pi/2$ - an overall phase-shift in the second path.
\item[(c)] $\phi_{s,p}^{2}=\pi/2$ and $\phi_{p,d}^{1}=\pi/4$ - an overall phase-shift in the second path, plus a phase-shift of the $d$-wave for channel 1.
\end{description}

Physically, intra- and inter-channel magnitude and phase differences of the partial-wave components are expected purely from the energy-dependence of the ionization dynamics. Contributions from the harmonic phase, or from other physical processes such as resonances at the 1-photon level in specific channels, can also play a role. Depending on the physical origin, such phase effects might shift all partial waves in a given channel (the simplest case of an optical phase shift in the XUV), or affect the photoionization dynamics in more complex and subtle ways. For further discussion on and around this point see, for example: refs. \cite{Seideman_1998} and \cite{Fiss_2000} for a general discussion and observation of resonant phase effects in photoionization, ref. \cite{Swoboda_2010} for a similar observation in RABBIT measurements, and ref. \cite{Jim_nez_Gal_n_2014b} for the case of autoionizing resonances in RABBIT-type measurements (recently demonstrated experimentally \cite{Gruson_2016}); ref. \cite{Yin_1992} for the related case of control over multi-path ionization schemes, specifically with $l$-wave parity breaking due to interfering 1 and 2-photon pathways, and ref. \cite{Laurent_2012} for application in an AR-RABBIT type experiment (see also sects. \ref{sec:even-harmonics} and \ref{sec:even-harmonics-ne} herein); ref. \cite{Wollenhaupt_2005} discusses conceptually similar cases of time-domain control schemes in photoionization, including temporal and polarization control in multi-photon ionization schemes.

For the usual sidebands, the amplitude of the resultant wavefunction will oscillate at $2\omega_{ir}$, with a total phase defined by the interfering partial-waves for each channel (including any contribution from the XUV optical phase). Within the approximations described above, the angular form of the sidebands will not show any time-dependence, since this requires a change in the relative phases of the contributing paths as a function of time. In this simple case, there are no dynamics which affect these quantities, and it is only the absolute amplitudes which vary as the IR-laser field oscillates. Hence, the angle-resolved interferograms will appear to simply breathe (in intensity) as a function of time. However, the presence of any time-dependence to the dipole matrix elements - e.g. Stark shifts affecting the ionizing states as the IR field cycles - would create additional time-dependence in the angular content, and might be expected in the strong-field regime.

Thus, in the usual regime, although the shape of the angular distribution is sensitive to the relative phases of the matrix elements, it is time-invariant; the total photoelectron yields are, however, sensitive to both the phases of the matrix elements and the instantaneous laser field. In particular, the phase shift of the yields relative to the laser field is sensitive to both the relative phases of the channels (hence may be used to probe the effect of resonances in one channel, as per ref. \cite{Swoboda_2010}), and the partial-wave phases within each channel. In this manner, the angular information provides a phase-sensitivity which is otherwise lost in an angle-integrated measurement.

\subsection{Sidebands in AR-RABBIT with non-linearly polarized light}

As illustrated in fig. \ref{fig:pathways}, the use of polarization states other than linear (and a parallel polarization geometry), will result in population of different $m$ states. In the most general case, where the XUV and IR fields have different polarization states, many additional pathways may play a role. Here a simplified case is illustrated, in which the XUV and IR fields are assumed to have the same ellipticity $\xi$, 
in order to illustrate the general concepts and trends with polarization state.

\begin{figure*}[h!]
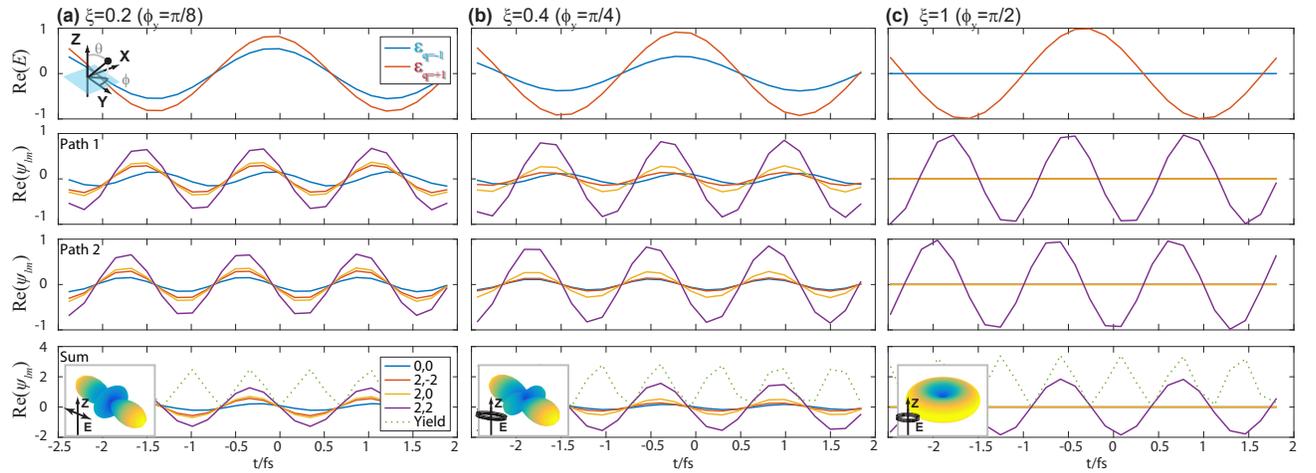

\begin{center}
\includegraphics[width=2\columnwidth]{{{basic_model_3up_m0_only_inc_PADs_polStates_080317}}}
\caption{{\label{fig:2-state_model_pol}Partial-wave resolved RABBIT traces for (a), (b) elliptically and (c) circularly polarized $E$-fields. Each column shows, from top to bottom, the IR laser field $E_{ir}^q$, the two paths to the side-band and the resultant sum over paths. The plots for each path show the real part of the final-state wavefunction $Re[\psi_{l,m}^c]$ as a function of partial-wave and time, and the sum over paths $Re[\sum_c\psi_{l,m}^c]$, where $c$ is a channel index for path 1 and path 2. The legends show the various $l,m$ partial-waves plotted, and the final (angle-integrated) photoelectron yields $I_2(\tau)$. Insets show the (time-invariant) angular interferograms, $I_2(\theta,\phi)$ summed over all channels. The reference geometry is illustrated in the first panel; the laser pulses propagate along the $z$-axis, and the ellipticity is defined in the $(x,y)$ plane. The fields are defined by ellipticity $\xi$, and (equivalently) the phase-shift of the $y$-component, $\phi_y$ (radians).%
}}
\end{center}
\end{figure*}

The results are shown in fig. \ref{fig:2-state_model_pol}. In these calculations, the model system detailed above is utilized, incorporating the set of phase shifts (c) (sect. \ref{sec:usual-RABBIT}). The three columns in the figure show the results for different ellipticities, defined mathematically by the phase shift between the two Cartesian components of the $E$-fields ($\phi_y$) (see ref. \cite{Hockett_2015} for details), and illustrated by the two spherical components of the IR field ($q=\pm 1$). The effect of the polarization state is quite clear here: as the polarization state moves from linear (equal magnitudes for the $q=\pm 1$ components) towards pure circular polarization ($q=+1$ in this example) the continuum wavefunction becomes increasingly dominated by the $|d,m=2\rangle$ component. In this (relatively) simple example, this is a direct consequence of the selection of pathway by the polarization state of the light: the handedness of the light is approximately mapped onto the $m=\pm 2$ final states. 
The most interesting case is, therefore, shown in  fig. \ref{fig:2-state_model_pol}(b), where the presence of both $q=\pm 1$ breaks the cylindrical symmetry of the distribution. 
In contrast, fig. \ref{fig:2-state_model_pol}(c), pure $q=+1$ light, produces a much simpler angular distribution, with only the $|d,m=2\rangle$ continuum state contributing. The symmetry breaking is also present in fig. \ref{fig:2-state_model_pol}(a), but is not yet pronounced with only a slight difference in the magnitudes of the $m=\pm 2$ states.

In this simple case, the additional pathways accessible with $q=\pm 1$ allow for breaking of the cylindrical symmetry of the angular distribution when the $E$-fields are elliptical, thus providing additional interferences, hence information content, in such measurements. Generally, the mapping between $q$ and the final observable is less direct, since many more states typically play a role. Examples for a more realistic case are given in sect. \ref{sec:Ne_elliptical_pol}. In traditional ionization studies, the use of polarization state and geometry is a powerful tool, and has been used in a variety of methodologies, for example in photoelectron metrology \cite{Reid_1991,Reid_1992,Leahy_1992} and control problems, including time-domain polarization-multiplexed schemes \cite{Hockett_2014,Hockett_2015c}. Recently, the related case of XUV field polarization effects on photoionization in the strong field regime has been investigated by Yuan, Bandrauk and co-workers (see, for example, refs. \cite{Yuan_2016, Yuan_2016a}); of particular note in that case is the presence of a strong radial (energy) dependence of the angular interferogram within a single photoelectron energy band, and asymmetries in the molecular frame. Polarization geometry in XUV-XUV 2-photon transitions have been investigated theoretically by the same authors \cite{Yuan_2017}, and XUV-IR schemes with polarization control have also recently been investigated experimentally \cite{D_sterer_2016}.

\subsection{Sidebands in extended AR-RABBIT with even harmonics \label{sec:even-harmonics}}

Additional interferences in the final state wavefunction can be created by adding ionization channels.  In a RABBIT experiment the addition of even-harmonics is the simplest scheme which achieves this, and is illustrated in Fig. \ref{fig:pathways}. Adding an interfering 1-photon channels has two effects: (1) time-dependence of the angular interferograms is now present, since the 1-photon channel is not coupled to the IR field, hence remains an invariant reference throughout the measurement; (2) the mixing of channels with odd and even photon order provides a route to parity breaking via the mixing of odd- and even-$l$ waves. 

In more detail, (1) implies that this scheme can be considered as a hetrodyne measurement, in which the 1-photon channel acts as a local reference for the 2-photon channels. This implies that additional information may be gained on the photoionization dynamics, since the usual sidebands are now additionally referenced to this 1-photon channel. In the usual case, the phase of the photoelectron yield provides relative phase information on the interfering 2-photon paths, referenced to the IR field. In this extended case, the overall phase remains referenced to the IR field, but the individual partial-wave phases play a more significant role in the time-dependence of the observed angular interferogram. In essence, one expects to see different features of the angular interferogram at different delays, and a much more complex time-dependence than the basic breathing mode of the usual RABBIT sidebands.

Generally, (2) applies to any scheme which mixes channels of odd and even photon order provides a route to parity breaking via the mixing of odd- and even-$l$ waves. While this type of final state control can be achieved in a number of ways (see, for example, refs. \cite{Yin_1992, Yin_1995, Wang_1996}), in a RABBIT experiment the addition of even-harmonics is the simplest and most appropriate route \cite{Laurent_2012}. In this specific case, one can view the temporal dependence of the resulting interferograms as a form of control, since this is nothing but a shift of the relative phase of the pathways defined by $\tau$; however, it is a relatively weak form of control, since the amplitudes of the 2-photon pathways are also dependent on the IR field. The use of additional $E$-fields, different polarization states, or shaped pulses, could all potentially provide more powerful means of interferogram control.

\begin{figure}[h!]
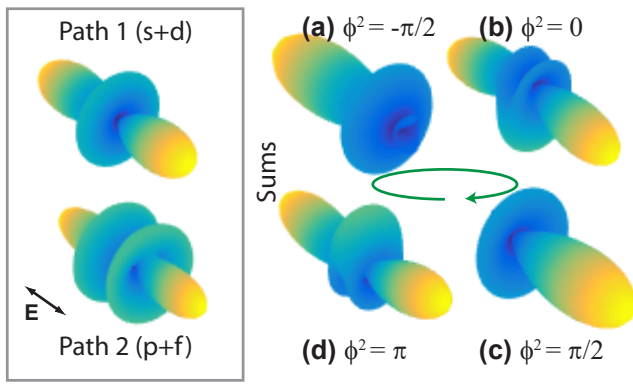

\begin{center}
\includegraphics[width=1\columnwidth]{{{model_PADs_parity_mix_080217b}}}
\caption{{\label{fig:odd-even} Basic concept of odd-even parity mixing in a two-path photoelectron interferometer. In this case, the usual cylindrical symmetry is broken, and the sum over paths varies as a function of the relative phase $\phi^2$.%
}}
\end{center}
\end{figure}

The basic concept of phase control is illustrated in Fig. \ref{fig:odd-even}, which shows the concept for a simplified two channel model. In this case, path 1 has only odd-$l$ components (as per the previous example, outlined in Sect. \ref{sec:usual-RABBIT}), and path 2 has only even-$l$ components. The phases of all components are set to zero, but a relative phase between the paths is varied in the model. The resultant wavefunction therefore takes the form $\Psi=\Psi_{1}+\Psi_{2}e^{i\phi^2}$. In this case, the change in the relative phase of the paths ($\phi^2$) results in different regions of constructive and destructive interference, with lobes in the final interferogram shifting as a function of phase. Again, this phase could be the result of the time-dependence of one path, as for (1) above, but could also be the result of another form of phase-control, or result from other dynamic effects. The full time-dependence of the angular interferograms in this class of scheme is discuss further in Sect. \ref{sec:real}.

\section{Real systems\label{sec:real}}

In order to treat real systems within the framework defined herein, a numerical treatment for the photoionization matrix elements (specifically, the radial integrals) for a given ionizing system is required. In this work, the bound-free matrix elements are computed using the ePolyScat suite \cite{Gianturco_1994,Natalense_1999,Lucchese2015}, and the continuum-continuum matrix elements treated as hydrogenic (similar to the treatment of ref. \cite{Dahlstr_m_2012}). This specific choice of numerical treatment is general, since ePolyScat is capable of accurate calculations for both atomic and molecular scattering systems, but is expected to be poor at low energies where the assumption of hydrogenic continuum-continuum transitions does not hold.

\subsection{Numerical details\label{sec:numerics}}
As discussed above, in order to model real systems numerical methods must be employed in order to determine the radial matrix elements (as distinct from the model cases above, in which the radial matrix elements are set as model parameters). In order to achieve this, a combination of numerical treatments was used:
\begin{itemize}
\item Bound-free matrix elements. For a given ionizing system and ionizing orbital, ePolyScat (ePS) can be used to compute dipole matrix elements. ePS takes electronic structure input from standard quantum chemistry codes, solves the continuum wavefunctions variationally with a Lipmann-Schwinger approach, and computes dipole integrals based on these wavefunctions; for further details, see refs. \cite{Gianturco_1994,Natalense_1999,Lucchese2015}.
\item Continuum-continuum matrix elements. Absorption of an IR photon in the continuum is modelled using Coulomb functions, in a similar manner to ref. \cite{Dahlstr_m_2012}, see sec. \ref{sub:CCcoupling} for details.  
\item VMI measurements.  To model the experimental VMI measurements, the input harmonic spectrum (800~nm driving field) was estimated as a series of Gaussians. Photoelectron energies then follow from the photon energies so defined, and IP of the ionizing system. This procedure also provided specific photoelectron energy points for the ePS and continuum-continuum calculations, and the matrix elements were assumed to be constant over the width of the spectral features and as a function of the laser field intensity.  See sect. \ref{sub:VMIdetails} for details.
\end{itemize}

In this manner, ionization of any given system, at a given photon energy, can be accurately computed (ePS), while the continuum-continuum coupling is approximated assuming Coulombic (asymptotic) continuum wavefunctions.

\subsubsection{Continuum-continuum coupling with Coulomb wavefunctions\label{sub:CCcoupling}}

The continuum wavefunctions in this case are, as previously (eqn. \ref{eq:Psi_gen}), given by a general expansion, which can be written in radial and angular functions. For the Coulombic case this is usually given as (see, e.g., ref. \cite{messiah}):
\begin{equation}
\psi_{lm}(\mathbf{k},\mathbf{r})=\phi_{l}(k,r)Y_{lm}(\theta,\phi)=A_{l}(k,r)F_{l}(r)Y_{lm}(\theta,\phi)\label{eq:Psi_genC}
\end{equation}

Where, 

\begin{equation}
A_{l}=\frac{2l+1}{kr}i^{l}e^{i\sigma_{l}}
\end{equation}

\begin{equation}
\sigma_{l}=\arg\Gamma\left[l+1-i\frac{Z_{1}Z_{2}}{k}\right]\label{eq:coulomb-phase}
\end{equation}

Here $F_{l}$ is a regular Coulomb function \cite{Abramowitz1970}, $\sigma_{l}$ is the (Coulomb) scattering phase, $Z_1$ and $Z_2$ are the charges on the scattering centre and scattered particle, and $\Gamma$ is the gamma function. Solutions of these equations can be computed numerically, as herein; analytical approximations have also been derived \cite{Dahlstr_m_2012}.

The explicit form of the continuum-continuum radial matrix element, for specific initial and final states defined by $|k,l,m\rangle$ is then given by:
\begin{equation}
R_{l_{i}l_{f}}(k_{i},k_{f})=\intop_{r}dr\,\phi_{l_{f}}(k_{f},r).r.\phi_{l_{i}}(k_{i},r)
\end{equation}

Of note in this case is the assumption of an $m$-independence to the scattering problem, which is correct over all $r$ for a Coulombic scatterer (point charge), but only correct asymptotically in general: hence this continuum-continuum form is appropriate only for overlap integrals at long-range from the ionic core in general.  For general discussion on short and long-range scattering, see ref. \cite{Park_1996}; for discussion of far-field onset in multipolar systems see ref. \cite{Knox_2010}; for discussion in the context of RABBIT see ref. \cite{Dahlstr_m_2012}. Physically, the characteristic ranges of the problem will depend on the scattering system and the precise details of the potential (which may additionally be affected by the IR field in cases of moderate to strong fields), and may need to be evaluated for specific cases when a high degree of accuracy is sought.

Finally, it is interesting to note that Dahlstr\selectlanguage{ngerman}ö\selectlanguage{english}m et. al. \cite{Dahlstr_m_2012} analyse these matrix elements analytically, and derive some approximate forms. Of particular interest is that the phase contribution from the continuum-continuum transition can be approximated as:
\begin{widetext}
\begin{equation}
\phi_{cc}(k_{i},k_{f})\equiv\arg\left\{ \frac{(2k_{f})^{iZ/k_{f}}}{(2k_{i})^{iZ/k_{i}}}\frac{\Gamma[2+iZ(1/k_{f}-1/k_{i})]+\gamma(k_{i},k_{f})}{(k_{f}-k_{i})^{iZ(1/k_{f}-1/k_{i})}}\right\} 
\end{equation}
\begin{equation}
\gamma(k_{i},k_{f})=iZ\frac{(k_{f}-k_{i})(k_{f}^{2}-k_{i}^{2})}{2k_{f}^{2}k_{i}^{2}}\Gamma[1+iZ(1/k_{f}-1/k_{i})]
\end{equation}
\end{widetext}
Here $Z$ is the nuclear charge, and the term $\gamma(k_{i},k_{f})$ is a long-range amplitude correction. This form, according to ref. \cite{Dahlstr_m_2012}, ``leads to an excellent agreement with the exact calculation at high energies". However, this comparison with exact results also indicated that it is not expected to work well at low energies, $<$8eV. 
Also of note in this approximation is that the continuum-continuum transition simply defines an energy-dependent phase-shift, with no $l$-dependence.


\subsubsection{Velocity Map Image (VMI) Simulation\label{sub:VMIdetails}}
\begin{figure}[h!]
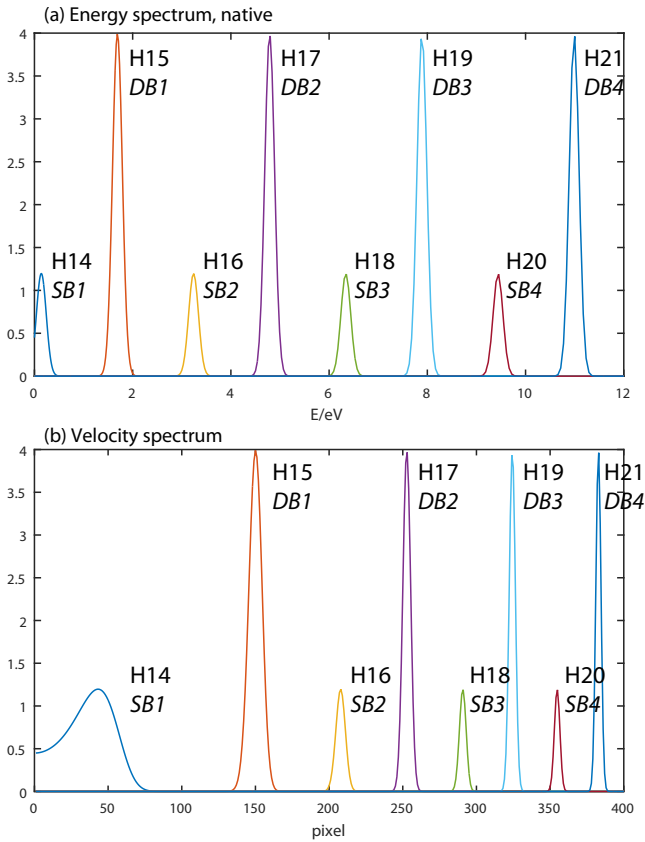

\begin{center}
\includegraphics[width=1\columnwidth]{{{VMI_E-scaling_0-12eV_800px_gw0.1_150216_edit}}}
\caption{{\label{fig:VMI-mapping}Example of energy to velocity space conversion in VMI, based on an approximate photoelectron spectrum. The main features correspond to direct ionization (direct bands, `DB'), reflecting the spectrum of the incident (odd) harmonics (Hn), while the minor features indicate sideband positions (`SB') in ``usual" RABBIT schemes, and also correspond to the position of even-harmonic orders. In this case, an 800~nm driving field was assumed, and an IP of 21.56~eV (1st IP of neon). Note that low-energy features are spread out in velocity space, hence appear as large central features in the final image, while high-energy features will be compressed, and appear as sharp outer rings in the image.%
}}
\end{center}
\end{figure}

In order to provide visceral results, and provide a more direct comparison with experimental measurements, the calculated photoelectron interferograms can be used to simulate velocity map imaging (VMI) measurements of photoelectron interferograms. 
Numerically, this involves calculating a volumetric (3D) space, simulating the photoelectron distribution and summing to form 2D image planes: full details of the approach can be found in ref. \cite{Hockett_2015}.  In the current model the radial (energy) spectrum is not calculated directly, so measured or estimated harmonic spectra are used to determine a set of Gaussian radial functions $G(k)$, as discussed above, which are then mapped to velocity space and used to describe each band in the measured photoelectron spectrum. An example is given in figure \ref{fig:VMI-mapping}, where the main features correspond to direct 1-photon ionization (labelled as `DB') by the input harmoic spectrum (odd-harmonics from an 800~nm driving field), and the minor bands correspond to the position of the 2-photon RABBIT sidebands (labelled as `SB') and even-harmonics (if present). These radial distributions are combined with the modelled angular distributions to determine the final photoelectron distribution on a 200x200x200 voxel array, and consequent 2D projections on a 200x200 pixel grid.


\subsection{AR-RABBIT results}
Following the prescription of sec. \ref{sec:numerics}, model results for photoionization of neon, and RABBIT measurements, were calculated. In modelling this case, ionization from a single initial state $|p,m=0\rangle$ was assumed for simplicity, corresponding to one component of the $2p$ valence orbital. 
Experimentally, one would assume that all degenerate $m$ states contribute equally, however the general phenonmenology and form of the results is unchanged by incorporating the degenerate $m=\pm 1$ initial states. 
Physically, the choice of a single $m$ state corresponds to a choice of reference frame and, potentially, a form of alignment: in the atomic case this can be considered as orbital polarization, while in the molecular case may correspond to an aligned molecular ensemble, or to the molecular frame \cite{Dill_1976}. 
The calculated photoionization matrix elements for the 1 and 2 photon transitions are given in the appendix (sect. \ref{sec:matElements}).

The position of the direct and sidebands calculated follow those shown in fig. \ref{fig:VMI-mapping}, which assumes an 800~nm driving field and the 1st ionization energy of neon (\href{http://physics.nist.gov/PhysRefData/Handbook/Tables/neontable1.htm}{21.56~eV} \cite{Kaufman_1972, NIST}). The lowest energy feature, SB1, is not accurately modelled in this case, since the $\Psi_+$ pathway corresponds to direct (and possibly resonant) 2-photon ionization, which is not defined by the simple 2-step model. (For discussion of the similar case of RABBIT measurements in He, which also involved a resonant channel, see ref. \cite{Swoboda_2010}.) However, this pathway was approximated by using the lowest energy bound-free matrix elements, and is included here to emphasize the velocity mapping effect, which causes this central feature to perceptually dominate the final VMI measurements. All other direct and sidebands are expected to be within the range of applicability of the model, although accuracy of the model is expected to vary slightly as a function of energy due to the form of the continuum-continuum matrix elements assumed.

\begin{figure}[h!]
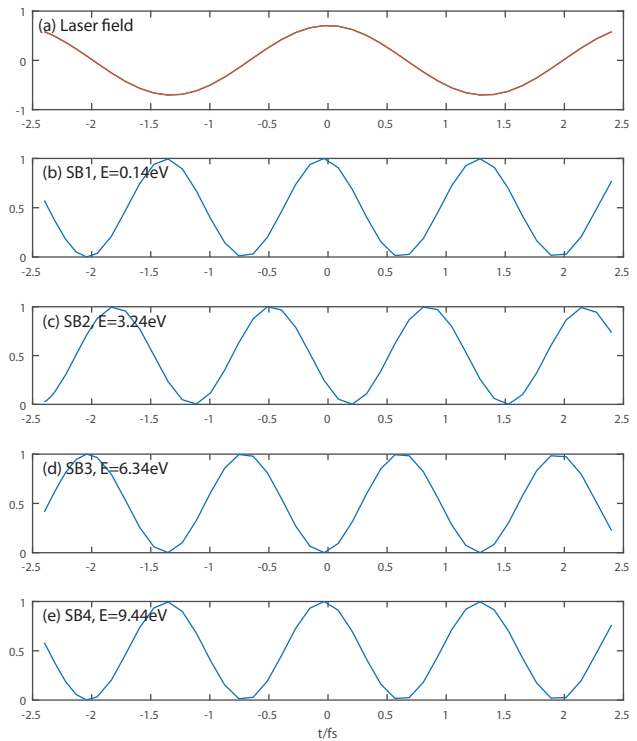

\begin{center}
\includegraphics[width=1\columnwidth]{{{Ne_RABBIT_yields_130217}}}
\caption{{\label{fig:Ne_1-colour_yields}Photoelectron yields (angle-integrated) for the RABBIT sidebands, $I_2({\tau})$, based on neon photoionization calculations. Each sideband is normalised to unity at the maxima.%
}}
\end{center}
\end{figure}

\begin{figure}[h!]
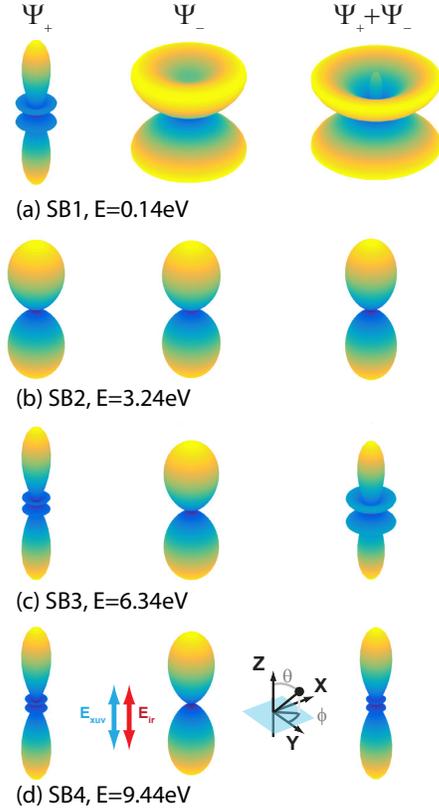

\begin{center}
\includegraphics[width=0.7\columnwidth]{{{Ne_1-colour_PADs_z-pol_3x4_100317b}}}
\caption{{\label{fig:Ne_1-colour_PADs}Photoelectron angular distributions for the RABBIT sidebands, based on neon photoionization calculations. In each case, the two contributing paths, $|\Psi_{\pm}|$, are shown as well as the resultant (channel-summed) interferogram.%
}}
\end{center}
\end{figure}

\begin{figure}[h!]
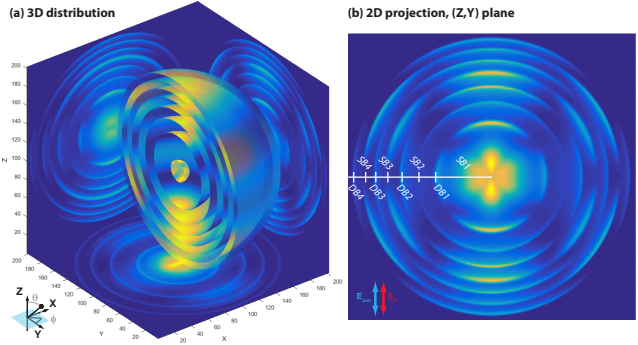

\begin{center}
\includegraphics[width=1\columnwidth]{{{VMI-sim_example_130217}}}
\caption{{\label{fig:Ne_1-colour_VMI}(a) Full 3D velocity distributions and simulated VMI measurements (2D projections), based on neon photoionization calculations, and incorporating the harmonic spectrum (odd-harmonics only) shown in fig. \ref{fig:VMI-mapping}. Each photoelectron band is shown by a single iso-velocity shell, colour mapped by the angular distribution. (b) Detail of the $(Z,Y)$ projection, with direct and sidebands labelled.%
}}
\end{center}
\end{figure}

Figs. \ref{fig:Ne_1-colour_yields} - \ref{fig:Ne_1-colour_VMI} provide a summary of the results. 
Fig. \ref{fig:Ne_1-colour_yields} provides the (angle-integrated) photoelectron yeilds, $I_2(\tau)$ for the four sidebands, and the corresponding, time-invariant, angular interferograms are shown in fig. \ref{fig:Ne_1-colour_PADs} for both contributing channels, and the resultant (channel-summed) observable. Fig. \ref{fig:Ne_1-colour_VMI} illustrates a set of iso-velocity (Newton) spheres from the full 3D photoelectron distribution, which each sphere corresponding to one band in the photoelectron spectrum, and the 2D projections of the full distribution.

A number of features are of note from these results:
\begin{enumerate}
\item As expected, the sideband phases vary according to the ionization dynamics (as a function of energy), incorporating both the direct ionization phase and the continuum-continuum phase.
\item The angular interferograms reflect the changing magnitude and phases of the $\Psi_+$ and $\Psi_-$ channels, and this is particularly apparent for SB3 and SB4. In these cases, it is primarily the relative phase of the SBs which contributes to the change in the final observable. The PADs change form significantly, and the temporal traces show a phase difference of approximately $\pi/2$.
\item The resultant PADs indicate structures with $L$ higher than the usual symmetry-imposed laboratory frame (LF) limit (for an isotropic initial state distribution) of $L\leq 2N$, where $N$ is the photon-order of the process \cite{Yang_1948,Dill_1976,Chandra_1987,Reid_2003}. However, these structures only follow from the assumption of polarized orbitals ($m=0$ selection), which allow a specific definition of the frame of reference. Additional $m$-state averaging over all initial $|pm \rangle$ components would reinstate the usual symmetry restriction; conversely, the presence of these structures in experimental measurements would provide evidence for orbital polarization, and this effect has recently been observed in AR-RABBIT measurements \cite{Niikura2015,Villeneuve2017}. As mentioned above, it is of note that these considerations are analogous to those for laboratory versus molecular frame measurements \cite{Dill_1976} and angular distributions from aligned molecules \cite{Hockett_2015b}.  
\item As discussed above, the simulated VMI measurements show a perceptual dominance of the lowest order bands due to the non-linear mapping from energy to velocity space, despite the fact that all bands are modelled in an identical fashion. Essentially, the energy resolution of VMI is non-uniform over the image, with the central region magnified relative to the outer region. This feature of VMI has previously been utilized to enable high-resolution spectroscopy \cite{Osterwalder_2004,Hockett_2009} and combined with field ionization for ``photoionization microscopy" experiments \cite{L_pine_2004}.
\end{enumerate}

Overall, these model results indicate some of the expected features of AR-RABBIT, as measured using VMI. Of particular note in this case is the fact that this modelling was motivated by recent work on neon AR-RABBIT measurements \cite{Niikura2015,Villeneuve2017}, in which aspects of the key features shown here were observed. In particular, the experimental measurements, performed at IR field intensities of $\sim10^{13}$~Wcm$^2$, revealed a 6-fold central structure, suggesting orbital polarization and selection in the strong laser field. It is, however, of note that this observation may also indicate higher-order photon processes than those expected ($N > 2$) contribute to the observable: in general careful intensity-dependence studies are required to determine which effect plays the key role \cite{Gal_n_2013}.

%
%

\subsection{Elliptically polarized light \label{sec:Ne_elliptical_pol}}

Following from the above, example AR-RABBIT results were also computed for an elliptically polarized IR field ($\xi=0.4$, as shown in fig. \ref{fig:2-state_model_pol}(b)), and a circularly polarized IR field. In these cases the XUV field was assumed to be linearly polarized, and a crossed polarization geometry was also assumed. In this geometry, again assuming a single initial $|p,m=0\rangle$ state, the XUV ionization accesses only $m=0$ states, while the IR field additionally accesses $m=\pm 1$ states. Essentially, this case allows for some, but limited, $m$-state mixing in the continuum-continuum transition. 

\begin{figure}[h!]
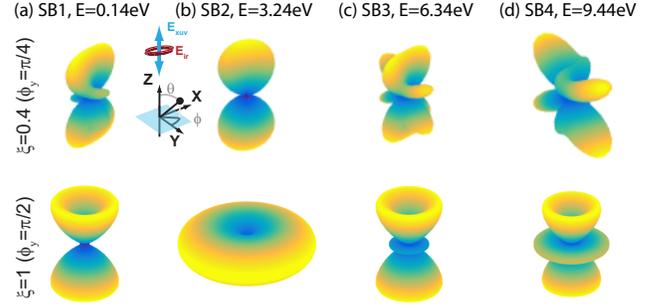

\begin{center}
\includegraphics[width=1\columnwidth]{{{Ne_2-colour_elliptical_pol_2up_100317}}}
\caption{{\label{fig:Ne_1-colour_PADs_pol}Photoelectron angular distributions for the RABBIT sidebands, based on neon photoionization calculations, for linearly polarized XUV and (top) an elliptically polarized IR pulse, (bottom) circularly polarized IR pulse, in a crossed polarization-geometry.%
}}
\end{center}
\end{figure}

Results are shown in fig. \ref{fig:Ne_1-colour_PADs_pol} for four sidebands. In the observables for the elliptically polarized case, the frame rotation between the XUV and IR field polarization vectors, and subsequent $m$-mixing in the continuum-continuum transition, results in ``twisted" structures (with specific handedness) appearing in the resultant distributions in most cases. It is of note that 2D VMI projections (fig. \ref{fig:Ne_1-colour_VMI}) will usually obscure such symmetry breaking, see e.g. ref. \cite{Hockett_2015a} and references therein for discussion; furthermore, other experimental factors which break spatial symmetry (e.g. a strong laser field) may also lift the $m$ state degeneracy in practice, and may thus constitute other mechanisms of spatial symmetry breaking. For the circularly polarized case, the lack of $m$-state interferences - since only a $m=+1$ states are accessed - results in a distinct, but cylindrical symmetric, distributions. Experiments utilizing this geometry are therefore particularly sensitive to any effects which break the $m$-state symmetry, such as a slight ellipticity in the XUV field or $m$-state mixing in a strong IR field.


\subsection{Extended AR-RABBIT \label{sec:even-harmonics-ne}}

Extended AR-RABBIT, in which even harmonics also contribute, 
presents the most information rich measurement. In this case, the interference between the time-dependent and time-independent channels provides an additional phase reference, creates interferences between channels with different photon orders, and results in a time-dependent angular interferogram. This provides the potential for control and metrology schemes analogous to many explored in previous energy-domain studies, such as odd-even parity mixing \cite{Yin_1992} and bound state resonance measurements \cite{Seideman_1998,Fiss_2000}: indeed, related concepts have already been explored in the time-domain \cite{Swoboda_2010,Laurent_2012,Laurent_2013}. In principle, it may also be possible to obtain a full set of partial wave magnitudes and phases using this technique (cf. ``complete" photoionization studies, e.g. refs. \cite{Duncanson_1976, Reid_1992, Reid_2003, Hockett_2014}) for a large number of partial waves, and the concept has recently been demonstrated for the atomic case \cite{Villeneuve2017}; equivalently, one can consider the technique as a means of obtaining full angle-resolved Wigner delays \cite{Dahlstr_m_2012,Hockett_2016}.

\begin{figure*}[h!]
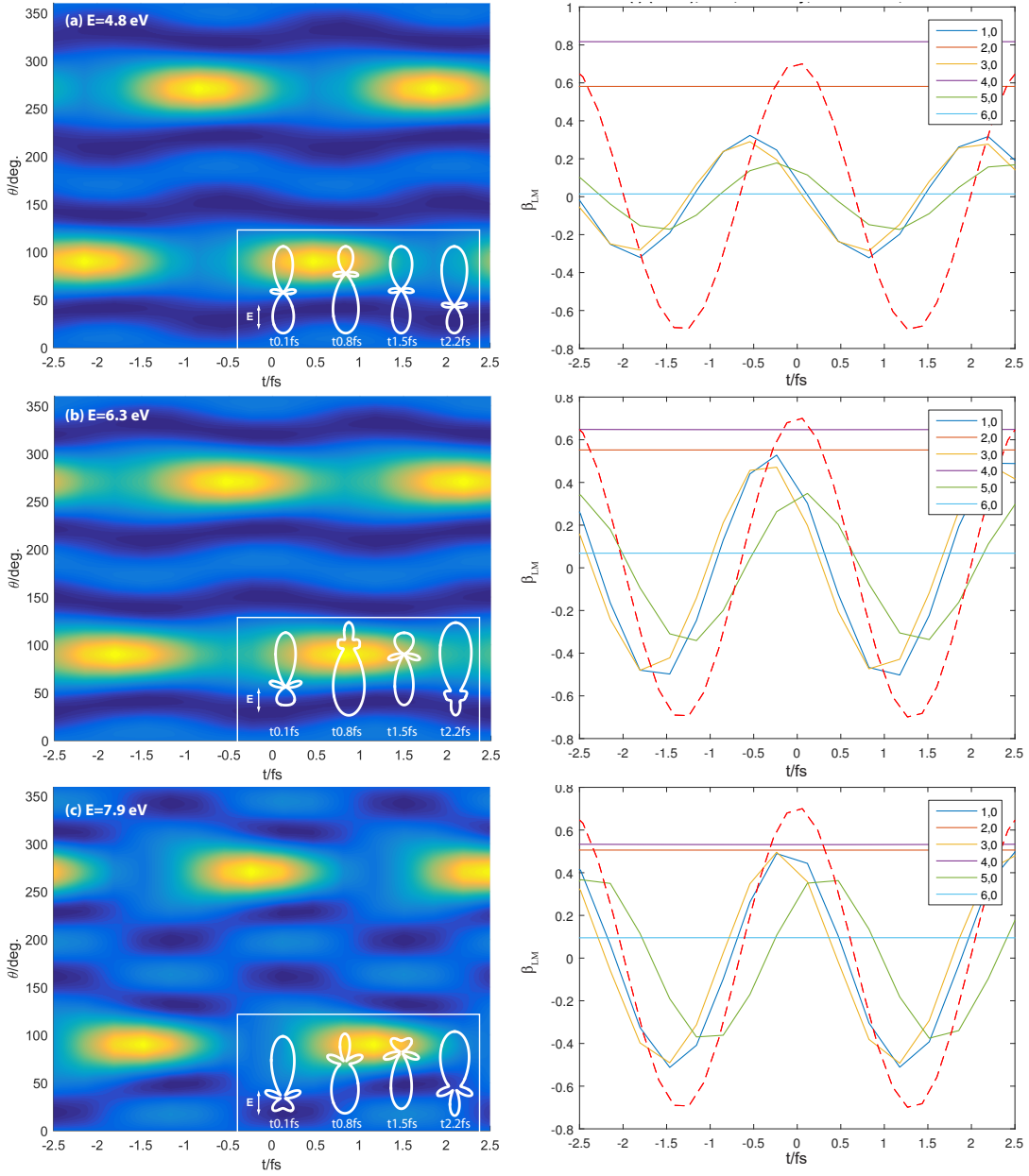

\begin{center}
\includegraphics[width=1.7\columnwidth]{{{Ne_2-colour_3up_layout_1D_PADs_010317}}}
\caption{{\label{fig:2-colour_observables}Time-dependent observables for three photoelectron bands for extended AR-RABBIT. Left column shows full $I(\theta,t; E)$ surfaces, with insets showing 1D cuts $I(\theta)$ in polar form at specific times. Right column shows the associated $\beta_{L,M}(t; E)$ values, normalised by the total yields $\beta_{0,0}(t; E)$.%
}}
\end{center}
\end{figure*}

The same model methodology as outlined above was employed, but with the addition of even harmonics in the XUV spectrum. 
Example results are shown in fig. \ref{fig:2-colour_observables}, which shows the resultant observables $I(\theta,\tau; E)$ and associated $\beta_{LM}(t; E)$ for three different photoelectron bands. In all cases complex behaviours can be observed, with multiple $l$-waves and phase contributions in the 3-path photoionization interfereometer leading to highly structured observables. 
Accross all of the bands, a similar structural motif is observed in the $I(\theta,\tau)$ plots, with the lobes along the laser polarization axis ($\theta=90,~270^o$) dominant, and weaker higher-order lobes. This structure is particularly clear in the polar plots given at discrete time-steps, and the corresponding $\beta_{L,M}(t)$ parameters, which contain both even and odd $L$ terms.

The time-dependence of the observables now contains contains two frequency components: even $L$ terms which oscillate at $2\omega$, and odd $L$ terms which oscillate at $\omega$. This basic behaviour has previously been observed and modeled by Laurent et. al. \cite{Laurent_2014}. 
However, the oscillation of the even terms corresponds to the same ``breathing" mode as described in the 1-colour case (since no additional cross-terms between the 1 and 2-photon pathways contribute), in which the photoelectron yield oscillated, but the angular distribution shows no time-dependence. 
Hence, normalisation of the angular interferograms by the total yield removes the time-dependence, and reveals time-invariant even $L$ terms. For this reason, no oscillations are observed in the even $L$ terms shown in fig. \ref{fig:2-colour_observables} (right column), and this part of the angular interferogram is time-invariant as for the ``usual" RABBIT case. The odd $L$ terms are more interesting, and result from the interferences between even and odd $l$-waves, correlated with the 1 and 2 photon transitions respectively. The effect of these interferences is, as noted above, to create up-down asymmetries in the observables. 
Clearly, the resultant interferograms are complicated, and the exact form of the observables are sensitive to the relative contributions and phases of the $l$-waves contributing to each of the three pathways. The relative phases observed in the $\beta_{LM}(t)$ can be considered as a probe of this behaviour, since different $l$-waves contribute to different $L$ terms \cite{Leahy_1991,Hockett_2015}; AR-RABBIT thus suggests a route to disentangling different phase contributions, related to the contributing ionization paths and $l$-waves, for use in phase-sensitive metrology scenarios. Of particular interest in this vein are ``complete" photoionization experiments, and angle-resolved Wigner delays, as noted previously.


Also noteworthy is the apparent temporal asymmetry of the observable in some cases: this is particularly apparent in the higher energy bands (e.g. band at 7.9~eV, fig. \ref{fig:2-colour_observables}(c)), with arrow-like structures spreading from the central lobes. This characteristic of the observable is a result of distinct temporal dependencies to the phases of the $l$-waves from different channels, leading to a skew in the temporal behaviour in some cases. Similar behaviour has previously been predicted based on a 2-path interferometer mediated by a vibronic wavepacket \cite{Hockett_2011}, which resulted in analogous $l$-wave intereferences; however,  in that case the asymmetry was not cleanly observed experimentally due the the temporal resolution of the measurement, although the results did strongly suggest such asymmetry was present. The presence of this type of temporal asymmetry in experimental measurements can therefore be regarded as a (relatively) direct phenomenological signature of significant phase-shifts between different $l$-waves. This characteristic is potentially useful as a means to observe experimentally-mediated changes in $l$-wave phases (e.g. due to laser intensity or wavelength) without the necessity of a full theoretical analysis of the results.

\section{Summary and conclusions}
In this work, some general properties of AR-RABBIT measurements have been investigated via a multi-channel, 2-photon ionization model in the perturbative regime. A range of interesting phenomena are observed in this case, due to the range of interferences contributing to the final observable. Of particular note is the fact that RABBIT type schemes mix bound-free matrix elements of different energies, which cannot be interfered in usual (energy-resolved) photoionization studies; furthermore, \textit{angle-resolved} RABBIT provides observables which are also highly sensitive to the $l$-wave amplitudes and phases (in direct analogy with traditional angle-resolved photoelectron spectroscopy). As discussed in the introduction, this presents AR-RABBIT as a potentially interesting methodology for any metrology schemes requiring phase-sensitivity to the ionization matrix elements as a function of angular-momentum and energy. Studies of photoionization dynamics in the energy and time-domain (Wigner delays) both come under this category, as does polarization-sensitive XUV pulse metrology. Experiments investigating the effects of bound-state or continuum resonances are one clear application of AR-RABBIT, and such effects can also be investigated as a function of the IR field intensity. The capabilities of ``extended" AR-RABBIT schemes, utilizing even haromics, are most interesting here, since 1 and 2-photon channels are interfered in this case, providing a hetrodyne type measurement, with the direct 1-photon channel as a time-independent phase reference. This scheme also allows for control over the resultant photoelectron interferogram, since up-down asymmetry can be broken as a function of IR field phase (i.e. XUV-IR time-delay).

Some of these concepts have already been investigated using RABBIT or AR-RABBIT techniques, but much work is open to fruitful exploration in this vein. Since VMI apparatus, along with other angle-resolved charged particle techniques (e.g. COLTRIMS), have proliferated in recent years, angle-resolved photoelectron measurements are now routine for many experimenters. This has lead to a range of novel studies utilising the related high-information content observable of photoelectron angular distributions \cite{Reid_2012}, and the outlook and utility of AR-RABBIT is similarly promising.

\section{Acknowledgements}
Special thanks to Hiromichi Niikura and David Villeneuve, for presentation and discussion of AR-RABBIT experimental results, which both suggested and motivated this study  \cite{Niikura2015, Villeneuve2017}. Thanks also to Ruaridh Forbes for suggesting the extension to elliptical polarization states, and Varun Makhija and Albert Stolow for general discussion.

\section{Appendix - Matrix Elements\label{sec:matElements}}
\begin{figure}[h!]
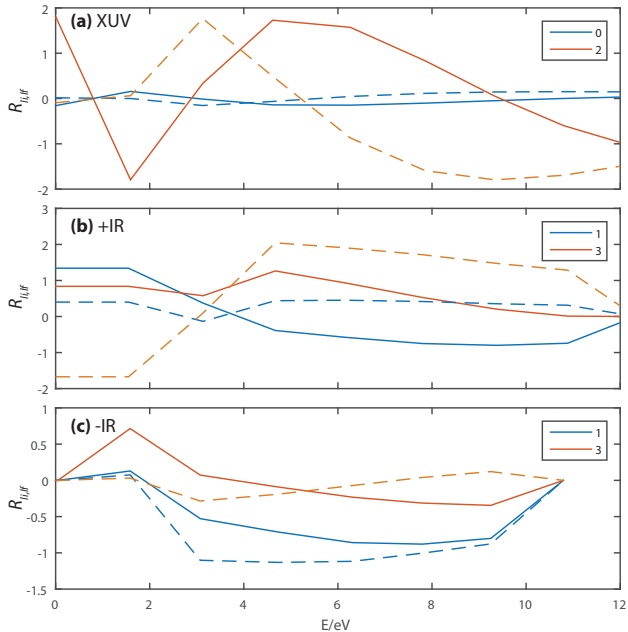

\begin{center}
\includegraphics[width=1\columnwidth]{{{matrix_elements_150317}}}
\caption{{\label{fig:mat_elements_Ne}Matrix elements for neon calculations. (a) 1-photon, XUV direct (ePolyScat calculations), (b) \& (c) IR continuum-continuum couplings, for absorption (+) and emission (-) channels respectively. In all cases, the matrix elements are indexed by final photoelectron states $($E$,l_f)$, and summed over multiple intermediate states $l_i$ where applicable. Solid lines show magnitudes, dashed lines phases. Final energy points match the photoelectron band centres.%
}}
\end{center}
\end{figure}

The full set of matrix elements for the neon calculations are shown in figure \ref{fig:mat_elements_Ne}. As detailed in sect. \ref{sec:numerics}, the 1-photon bound-free matrix elements were computed using ePolyScat, while the continuum-continuum elements assume Coulomb wavefunctions. In all cases the matrix elements are shown as a function of the final photoelectron energy. For the 2-photon bands the calculations assume an 800~nm IR field, hence $h\nu=1.55~eV$, and this is the energy difference assumed between the final and intermediate (1-photon) states in the calculation.

\bibliographystyle{apsrev4-1}
\bibliography{bibliography/converted_to_latex%
}

\end{document}